# Magneto-optical Kerr effect of a $Ni_{2.00}Mn_{1.16}Ga_{0.84}$ single crystal across austenite and intermartensite transitions

Jan Fikáček[1], Vít Kopecký[1], Jiří Kaštil[2,3], Oleg Heczko[1,3], and Jan Honolka[1]

[1]*Institute of Physics, Czech Academy of Sciences, Na Slovance 1999/2, 182 21 Prague 8, Czech Republic*

[2]*Institute of Physics, Czech Academy of Sciences, Cukrovarnická 10/112, 162 00 Prague 6, Czech Republic*

[3]*Faculty of Mathematics and Physics, Charles University, Ke Karlovu 3, 121 16 Prague 2, Czech Republic*

Abstract

We carried out magneto-optical Kerr effect (MOKE) and magnetization measurements on a single crystal of $Ni_{2.00}Mn_{1.16}Ga_{0.84}$ in a longitudinal geometry. While there had been reports of polar geometry measurements so far, we show that - against earlier predictions - surface magnetic states of the martensite and the austenite can be also probed efficiently via longitudinal MOKE. A single-variant magnetic state, which occurs at room temperature, is characterized by ferromagnetic hysteresis loops similar to standard ferromagnets. Temperature dependencies of Kerr rotation were found to be linearly proportional to magnetization for martensitic phases. After passing through an inter-martensitic structural transition at around 260 K in zero magnetic field, the coercive fields are more than doubled in comparison with the room temperature values. On the other hand, at higher temperatures above 338 K where an austenite structure is formed, the MOKE signals are dominated by quadratic contributions and the magnitude of Kerr rotation drops due to changes in the electronic and magnetic domains structure. After the sample is cooled across the austenite-martensite transition again, we observed a multiple beam reflected by the sample due to different types of structural domains present on its surface.

Email address of the corresponding author: fikacek@fzu.cz

1. Introduction

Ni$_2$MnGa alloys exhibit magnetic shape memory effects (MSM), which are manifested as extraordinary large anisotropic changes of the sample dimensions (up to 10 %) under an applied magnetic field [1–3]. The underlying mechanism is a rearrangement of ferromagnetic twin domains under the influence of magnetic fields [4]. These compounds falling into Heusler alloys [5] have been thoroughly studied for a long time, due to their potentials for future application, e.g. for actuator devices or for energy recuperation. Magnetic properties of a stoichiometric compound Ni$_2$MnGa were first studied by Webster et al. [6]. The average magnetic moment of 4.17 $\mu_B$ per formula unit is almost exclusively confined to Mn atoms, while only small moments can be found on Ni atoms (< 0.3 $\mu_B$). As a result of an indirect exchange interaction between Mn magnetic moments, the compound orders ferromagnetically below 376 K. Below 220 K, moreover, the cubic austenite structure transforms to a complex tetragonal structure via a martensitic transformation. It is of first order type and is accompanied by thermal hysteresis with coexistence of both phases. The transition is diffusionless and causes a variation in the Mn-Mn inter-atomic distances. For off-stoichiometric alloys during cooling, there is a cascade of further transitions between martensitic phases, so-called inter-martensite transitions (IMTs), having different structure modulations [4, 7] usually ending in a non-modulated tetragonal martensite [8]. It has been found that the Curie temperatures as well as the structure transformation temperatures can be strongly modified via small stoichiometry changes [9]. Related magnetic properties of a non-stoichiometric Ni$_2$MnGa [4, 10] and a stress dependence of the IMT were also studied [11]. During the IMT, the magnetization evolves smoothly unlike the martensite-austenite transition where it changes abruptly [4, 12].

Magneto-optically active materials offer to store information in the magnetic state of the material and the information can be instantly read optically via the optical activity. Materials with net magnetization within the light penetration depth have different refraction indexes for circularly left- and right-polarized light. As a result, the plane of polarization slightly rotates by the Kerr rotation angle ($K_r$) from its original direction and, when measured in reflection, produces the magneto-optical Kerr effect (MOKE). In most cases, only a linear dependence on magnetization $\sim M$ dominates MOKE signals. But there can be found examples where a quadratic contribution $\sim M^2$ plays an important role (Voigt effect). For in-plane magnetization vectors, they enter the signal as $M_L M_T$ [13] and $M_L^2 - M_T^2$ [14] terms, where $M_L$ and $M_T$ denote a longitudinal (parallel to the field and laying in the plane of incidence) and a transversal (perpendicular to the field and the plane of incidence, see Fig. 1a) component of magnetization, respectively. Studying surface magnetism via this method is straightforward and easy to realize using a linearly polarized laser beam, an analyzing polarizer and a detector of the passing light intensity. MOKE measurements are not limited only to ferromagnetic materials [15], but can also probe properties of superconductors [16]. Thanks to the fact that MOKE collects information only from a penetration depth of few nanometers (10-20 nm in metals [17]) one can also study magnetic multi-layers [18, 19], or topological insulators weakly coupled to ferromagnets [20].

Polar MOKE measurements on the specific $Ni_{2.00}Mn_{1.16}Ga_{0.84}$ compound have been recently performed by Veis et al. [21]. An emphasis was given on the spectral dependence of the MOKE signal at different temperatures. According to this, using a monochromatic laser with an energy of 1.85 eV, only a small MOKE signal should be visible, at least 10 times smaller in comparison with the whole energy range studied by Veis et al. In the austenitic phase the Kerr rotation is further weakened due to changes in the topology of the Fermi surface. In older literature it was even stated that longitudinal MOKE is negligible in this type of compounds [22].

The paper is structured in the following way. At first we show MOKE field scans below room temperature to investigate magnetic properties of different martensitic phases, namely their Kerr rotations and coercive fields, and discuss the data. Afterwards we focus on a high-temperature region where the martensite to austenite transition takes place and where a significantly different behavior has been found in the austenitic phase. MOKE results are compared to SQUID and electrical transport results on the very same sample. In the last two sections we discuss both temperature regions and make conclusions.

2. Experimental details

The single-crystal sample of dimensions 1.9x3.8x9.5 mm$^3$ and mass 0.502 g was cut via spark erosion from an ingot, which had been grown by a modified Bridgman method. The cuts were oriented perpendicularly to the main crystallographic directions ([100], [010], [001]) of the high-temperature austenite. Afterwards the sample was mechanically polished and subsequently electrochemically polished in order to remove the mechanically stressed layers of the surface. Our single-crystal had an approximate stoichiometry $Ni_{2.00}Mn_{1.16}Ga_{0.84}$ determined by EDX analysis. At room temperature, the sample structure corresponds to a 10 M modulated martensite [21] having a monoclinic (pseudo-tetragonal) crystal structure with lattice parameters $a_M \approx b_M = 5.93$ Å, $c_M = 5.6$ Å. As we will show below, its Curie temperature and the martensite-austenite transition all take place above room temperature.

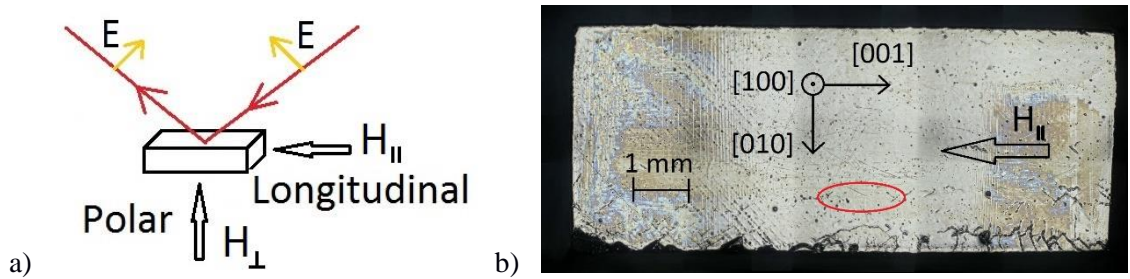

Fig. 1: a) A scheme showing arrangements of the MOKE setup in the longitudinal and the polar geometry (magnetic fields in- and out-of-plane, respectively). The yellow arrows denote the orientation of the light $E$-field vector, which is in the plane of incidence (p-polarization).

b) An optical microscopy image of the sample surface (~10 mm x 4 mm) with drawings indicating the main crystallographic axes orientations in a single variant state at room temperature. The field direction $H_{ll}$ parallel to [001] in longitudinal MOKE is shown as a large arrow and the red ellipse borders mark the area irradiated by the laser beam.

Electrical resistance was measured in a closed cycle refrigerator via the standard four probe method between 350 and 4 K (temperature sweep of 1K/min) with a current of 10 mA with switching polarity and running through 50 μm silver contacts attached to the sample surface by silver paste.

Magnetization at room temperature was measured by a vibrating sample magnetometer (VSM) under magnetic fields (up to 1.5 T) produced by an electromagnet. For this, the sample was fixed to the holder by wrapping it with a Teflon tape. Magnetization measurements in a temperature range going between 140 and 400 K with a heating/cooling rate of 2K/min were performed in a SQUID magnetometer and a Physical Property Measurement System (PPMS) equipped with a VSM option both made by Quantum design. In the first case the sample was fixed in a plastic straw by concentric pieces of another plastic straw and in the former by GE varnish.

The MOKE measurements were performed using a variable temperature magneto-optical setup consisting of a laser source (red color, 670 nm wavelength, energy 1.85 eV, power output less than 1 mW), a detection diode with a cooling system, and a set of three electromagnetic coils producing magnetic fields (two for an in-plane field and one for an out-of-plane field). An upper field limit for the former was 0.13 T and for the latter 0.08 T. In this work, we focus on the longitudinal geometry, with fields applied along the longest sample dimension, which is prepared in a single-variant state with the *c*-axis pointing in the same direction. This way the sample could be saturated even at low fields, since the *c*-axis is the easy magnetization direction and the demagnetization field is the lowest. The sample was glued to a copper plate by a GE varnish, which then was tightly screwed to the copper cold finger of an Oxford cryostat (MicrostatHe2, rectangular tail). The sample chamber was pumped to high vacuum ($10^{-6}$ mbar) and cooling of the cold finger was realized by a continuous liquid helium flow. Typical hysteresis curves were derived by an averaging over ten consecutive loops. Each point took 2 seconds including the collecting time (100 ms) and the time delay for magnetic fields to be stabilized. The signal usually exhibited a small time-dependent drift that was subtracted linearly via merging the first and the last point of the corresponding curve. Finally, each curve was normalized to 10 mV for its first zero-field point. For calculating of the Kerr rotation we took the difference between the intensities corresponding to the maximum negative and positive field. This difference was divided by the derivative of the calibration curve, which was determined via scanning the angular dependence of the intensity on the angle of the analyzer. The analyzer was usually set to an angle about 92 degrees from the polarizer angle to have a good sensitivity. We mostly used the longitudinal geometry (Fig. 1 a)) and p-polarized light. All MOKE data shown in this work were measured with these settings.

In order to systematically study how MOKE evolves during austenite-martensite and IMTs, we used the following measurement sequence, referred to as *experiment* #1 - #4. A first fast cooling experiment (*cooling #1*) served to verify a well-defined initial state of the sample and the observability of the IMT on the fly. After that, the respective range of IMTs were measured under thermal quasi-equilibrium conditions, which involved both slow cooling to low temperatures and slow heating back to room temperature (*cooling #2* and *heating #2*). In order to explore the martensite-austenite transition, *heating #3* experiment was done via warming up the sample above room temperature. The following experiment, *heating #4*, served for a comparison of Kerr rotations and coercivities among all studied structures. Thus it included heating from lowest temperatures up to 350 K, when the material is in the austenitic structure. And finally, the last experiment, *heating #5*, was a reproduction of *heating #3*.

3. Results

As an initial characterization, in Fig. 2 we show a comparison between magnetization measurements $M(\mu_0 H)$ taken by a VSM magnetometer and a longitudinal field scan by MOKE at room temperature. In order to get the sample into the single-variant state (Fig. 1 b)), where the *c*-axis points at the same direction within the whole volume, we magnetized the sample by a 1.5 T field applied along the longest dimension of the sample parallel to [001] in Fig. 1a) and 1b) (initialization procedure) before each of the MOKE experiments. If this procedure is not done correctly, the sample persists in a so-called multi-variant state in which the main axes orientations are not the same within the whole volume. After several thermal cycles, the sample had to be pressed along its longest dimension prior to the initialization procedure to achieve the single-variant state. This helped to preferentially set the shortest axis (*c*-axis) parallel to the pressed direction. Afterwards the magnetization curve of a single-variant-like type was restored independently of the sample history (the black short-dashed line in Fig. 2). At 1.5 T field, the magnetization was 70.3 Am$^2$/kg. The coercive fields found in the MOKE scans (see the inset of Fig. 2) are slightly bigger than those from bulk-sensitive VSM measurements most probably due to the influence of a surface pinning of magnetic domains [21]. In the multivariant state, the coercive force was even higher, which can be ascribed to a domain pinning on twin interfaces (boundaries).

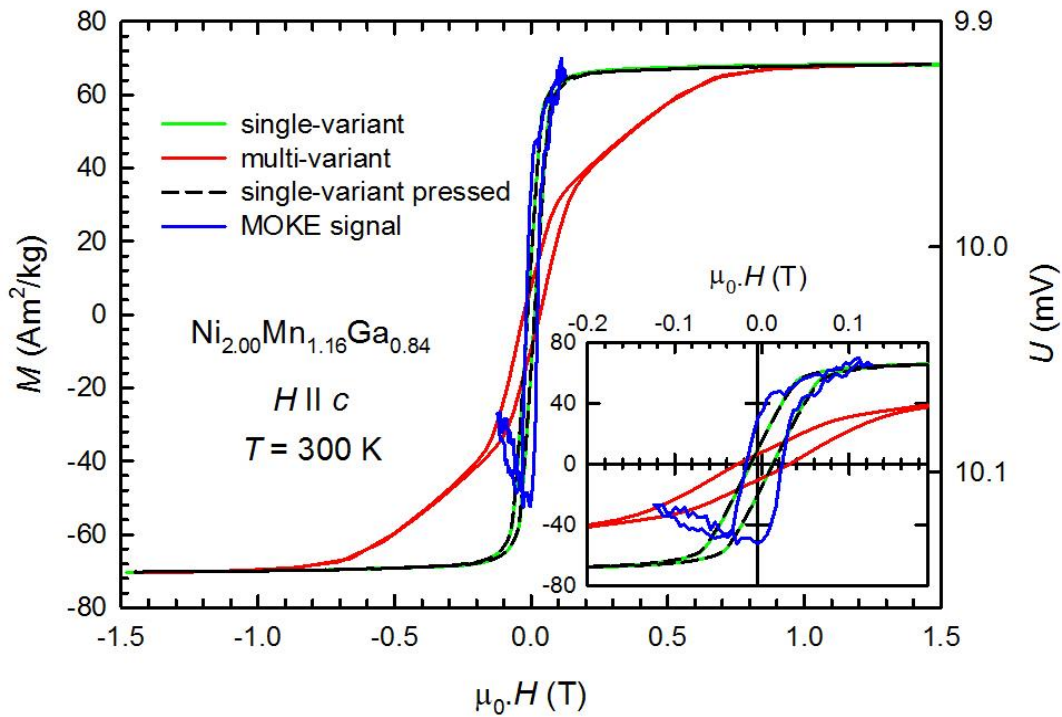

Fig. 2: Ferromagnetic hysteretic behavior of Ni$_{2.00}$Mn$_{1.16}$Ga$_{0.84}$ measured via VSM (left axes) and by MOKE (right axes) at room temperature. MOKE data corresponds to the sample in a single-variant state. For comparison, the VSM magnetization curves $M(\mu_0 H)$ are shown both in single-variant and multi-variant states. The inset shows the data in a low-field range. VSM data measured in a multi-variant state and after subsequent reinitialization are shown in red and black dashed line, respectively. We comment that the MOKE axes are inverted to have the same loop polarity as the VSM data.

To identify transition temperatures among different crystal structures and the Curie temperature, we took temperature scans of electrical resistance $R(T)$, and magnetization $M_{1T}(T)$ measured at $B = 1$ T, which are shown in Fig. 3 for a direct comparison. The electrical resistance exhibited a drop below room temperature revealing the IMT. The IMT is connected with a large thermal hysteresis, because it occurred at $T_{\text{IMT, cooling}} = 213$ K and $T_{\text{IMT, heating}} = 280$ K. Our data are similar to already published data [4, 10, 11]. A smaller step in $R(T)$ at around 170 K is a sign of another IMT to a non-modulated martensite as observed for example in [12]. On the other hand, $M_{1T}(T)$ shown in Fig. 3 varied more or less smoothly in the same temperature region in agreement with previous observations [4].

The electrical resistance measurements were extended down to 4 K (not shown). Using this data, we calculated the residual resistivity ratio (RRR) $R_{300K}/R_{4K} = 1.8$. Considering usual RRR values in metals, the value resembles polycrystalline samples. In our case, however, the lower RRR values most likely reflect disorder effects caused by its off-stoichiometric composition and a big number of twin boundaries (interfaces). A close examination of the electrical contacts on the sample after the experiment revealed signs of deteriorations, which are consistent with large changes of the sample surface dimensions and a relief during the martensitic transition. Above room temperature, a sharp step connected with a structural transformation into the austenitic phase is visible at $T = T_{\text{aust.}}$ in both electrical resistance and magnetization. The Curie temperature $T_C = 369$ K was determined as an inflection point in the temperature dependence of magnetization. There was a small temperature mismatch between electrical resistance and magnetization data most probably caused by poorer thermal contact in the former case. While the anomaly in $R(T)$ occurs at 344 K during heating and back to the martensitic structure at 318 K during cooling, in $M_{1T}(T)$, it takes place at 339 K and 326 K. The magnetization data were corrected for the cooling/heating rate. In addition, during the $M_{1T}(T)$ measurements the sample was being thermalized by surrounding He gas, but for the measurements of $R(T)$ it was in vacuum, so the thermal contact was poorer, and thus we can rely more on the precision of the determination of transition temperatures in $M_{1T}(T)$ than in $R(T)$. The determined values also agree well with the optical measurement shown later.

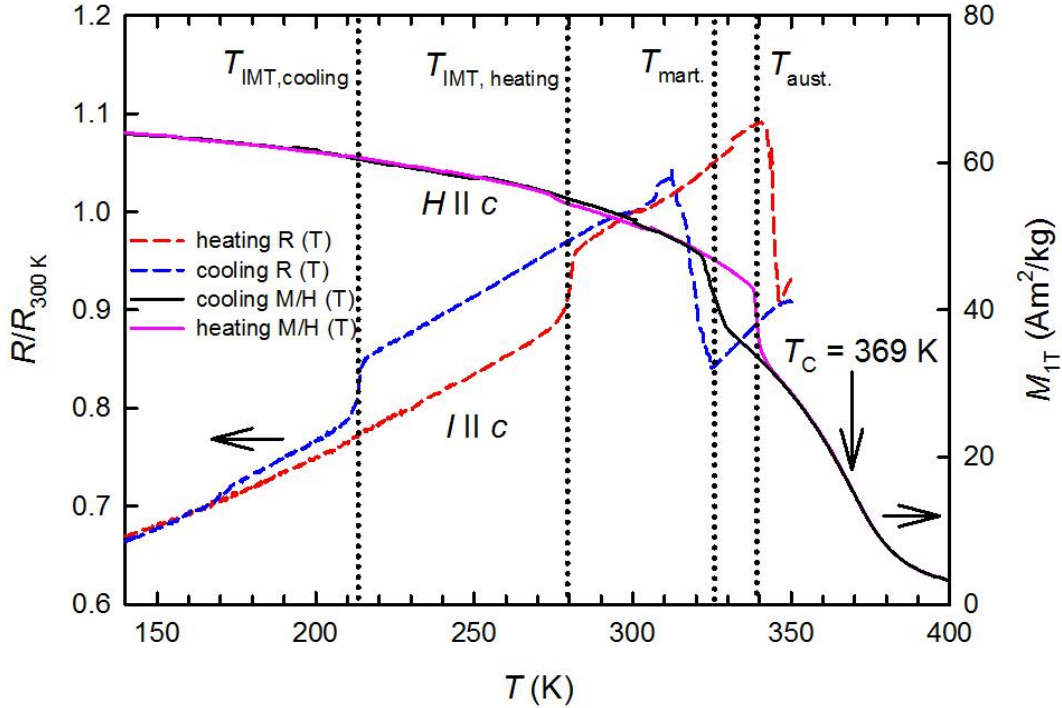

Fig. 3: Temperature scans of electrical resistance (dashed lines) and magnetization in a field of 1 T ($M_{1T}$) (full lines) both measured along the sample´s longest dimension (c-axis). Electrical resistance data were normalized to the room temperature values for a better comparison of different runs. Vertical dotted lines signify a formation of the martensite during cooling at $T_{mart.}$ = 326 K and the IMT at $T_{IMT,\,cooling}$ = 213 K as deduced from the steps in the magnetization and electrical resistance data, respectively. During heating, the IMT takes place at $T_{IMT,\,heating}$ = 280 K and the austenite appears above 339 K. A vertical arrow marks the Curie temperature $T_C$ = 369 K taken as the inflection point of the $M_{1T}$ temperature dependence.

Temperature dependent MOKE field scans were then performed during cooling from room temperature to about 180 K as shown in Fig. 4 (*cooling #2* and *heating #2*). The temperature dependency of coercivity and Kerr rotation obtained from these data are plotted in Fig. 5 together with data of *cooling #1*. Between 220 K and 200 K, where the IMT is expected, the coercive field increased steeply by a factor of two from 0.023 T to 0.054 T. On the other hand, coercive fields determined from bulk magnetization data from VSM and SQUID display only a modest increase with cooling. Contrary to this, the respective Kerr rotations did not change abruptly but varied smoothly in the same temperature range concurrently with the magnetization (Fig. 2). During subsequent heating to room temperature, the coercive field values from MOKE data decreased again between 260 K and 300 K, which agrees with the IMT temperature $T_{\text{IMT, heating}} = 280K$ determined by the electrical resistance measurements. *Cooling #1* experiment shown in Fig. 5 involved fewer temperatures, but the data coincide well with the second experiment within their error bars. Differences in magnitudes of Kerr effect originated most likely from surface inhomogeneity and limited reproducibility of MOKE measurements.

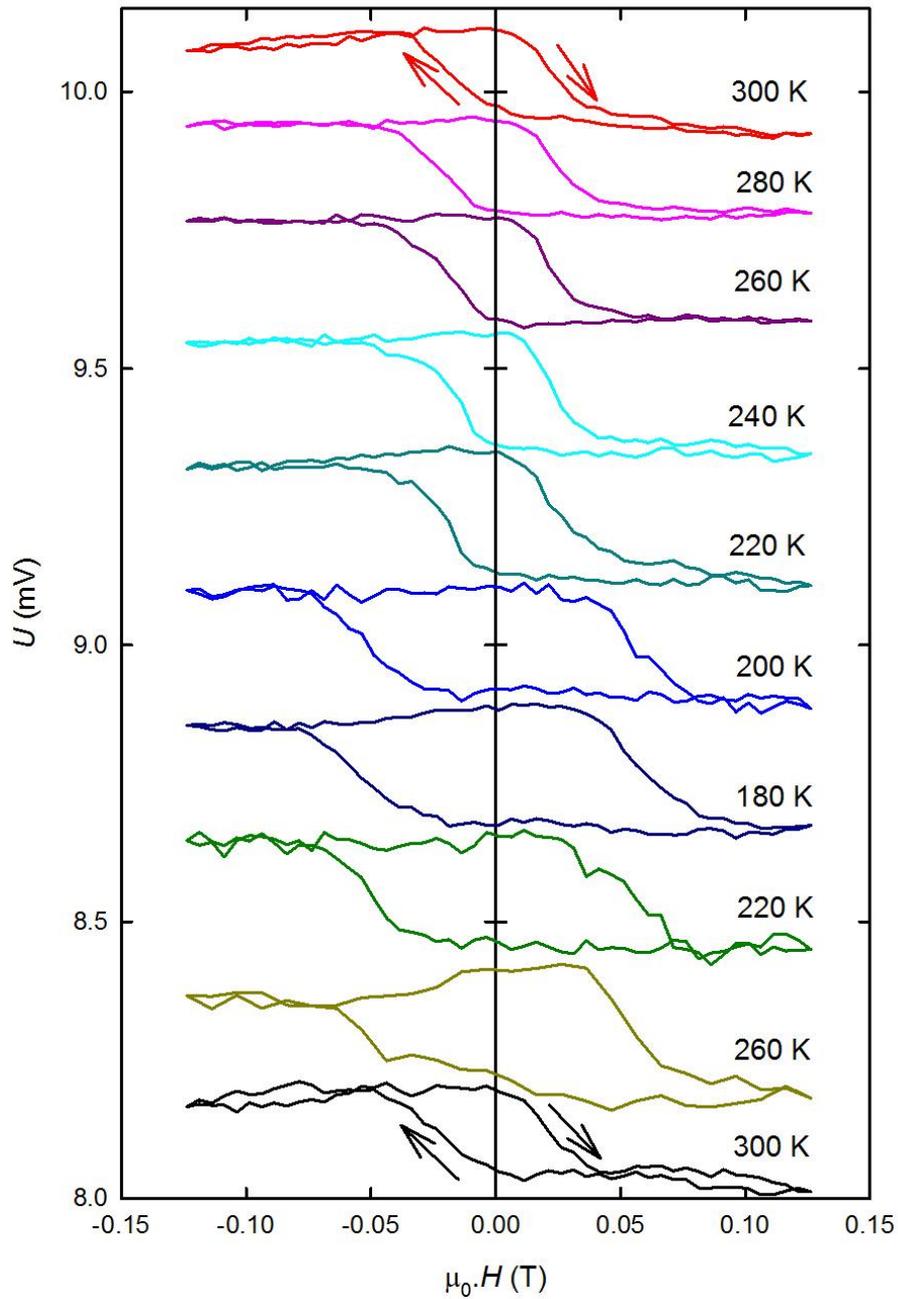

Fig. 4: MOKE field scans taken on single-variant states of the martensites during *cooling #2* and *heating #2*. The uppermost curve was taken in the initial state and the other plots are ordered downwards reflecting the timeline of the cooling procedure. The curves are shifted vertically for clarity.

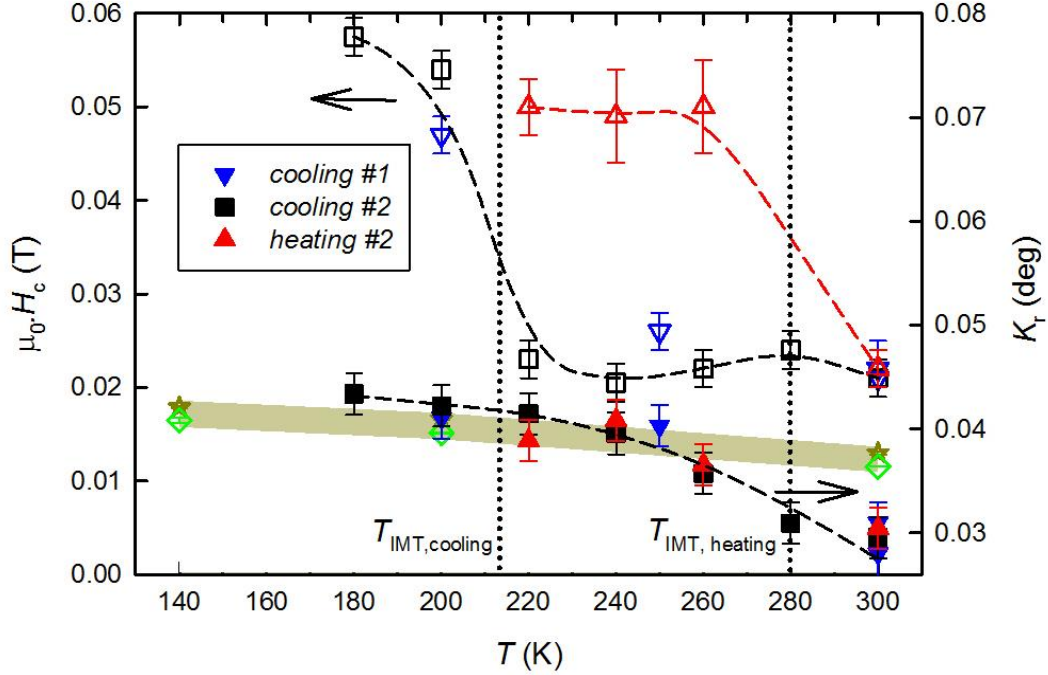

Fig. 5: Temperature dependencies of Kerr rotation ($K_r$, full symbols), and of coercive fields ($\mu_0 \cdot H_c$, empty symbols) calculated from the plots in Fig. 4 and from magnetization data, VSM (green diamonds) and SQUID (yellow stars) measurements. The semi-transparent stripe signifies a range of coercive fields estimated from the bulk data within the experimental error. Two vertical dotted lines mark the IMT temperatures determined from electrical resistance measurements. *Cooling #1* was not displayed in Fig. 4. In the second experiment, the data were collected during cooling and heating of the sample. Please note the different scales for $K_r$ and $H_c$. The dashed lines are just guides to the eye. Whilst the Kerr rotations differ between heating and cooling only in the order of the experimental error, the coercive fields (from MOKE) exhibit a large thermal hysteresis ≈ 70 K.

In the following, we want to discuss MOKE results obtained in a higher temperature range above room temperature (*heating* #3). Crossing the transition temperature from the martensitic to the austenitic phase at $T_{aust}$ = 338 K, the hysteresis loops change drastically towards wing-like loops at $T$ = 340 K (Fig. 6). As a consequence, there is a small region around zero field where the field up and field down curves overlap making it impossible to determine coercive fields. This wing-like shape was observed in several repetitions of the experiment later. At $T$ = 360 K, approaching the Curie temperature $T_c$ = 369 K, the wing-like characteristics disappears again and leaves a standard s-shaped curve typical for soft magnetic materials with almost zero coercivity. Going to a higher temperature of $T$ = 380 K in *heating #4*, the sample was in its paramagnetic state resulting in a negligible Kerr rotation signal below the detection limit.

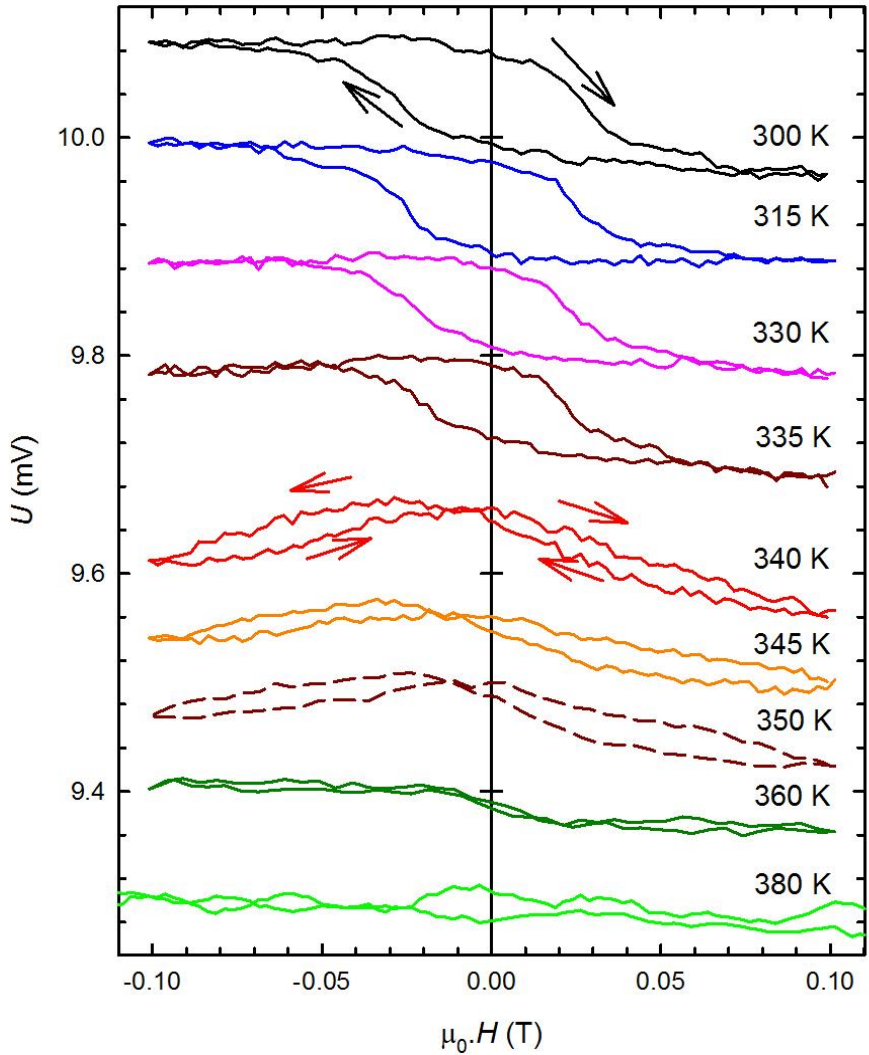

Fig. 6: MOKE field scans collected during *heating #3* above room temperature reflecting the transformation to the austenite state at $T_{aust}$ = 338 K. The upper curve was taken in the initial state and the consecutive measurements at increasing temperatures are ordered chronologically downwards. The last curve at $T$ = 380 K was obtained in *heating #4*. All curves are normalized and shifted for clarity.

In Fig. 7 we summarize Kerr effect measurements above room temperature up to the paramagnetic state. There is a clear drop in the magnitude of Kerr rotation when the structure transforms into the austenite phase at $T_{aust}$.

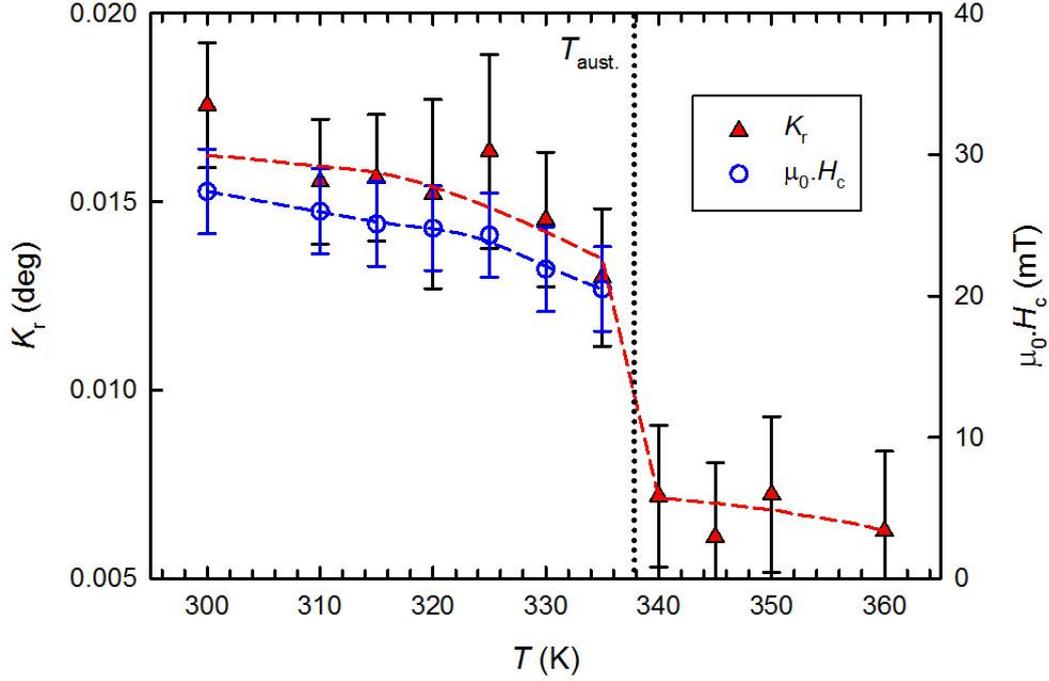

Fig. 7: A summary of the coercive fields and Kerr rotations assessed from the data in Fig. 6 during *heating #3*. The dashed lines are guides to the eye. At $T_{aust} = 338$ K a transition to the austenite state is marked by a vertical dotted line. Above it, due to the wing-like shape of the curves, the coercive fields are not defined in a standard way anymore.

In order to gain further understanding of the intimate connection between the rich structural phase diagram of $Ni_2MnGa$ compounds and magneto-optical properties, it is instructive to reenter the martensitic phase via cooling without external magnetic field. In this case, during cooling across the ferromagnetic ordering temperature, the microstructure of martensite changed to a twinned multivariant state, i.e containing large amount of differently oriented twin domains in order to secure elastic compatibility and to minimize the internal stresses. Contrary to the initial single-variant state enforced by the initialization procedure with 1.5 T bias fields parallel to [001], a multi-variant state was established during zero field cooling. Using optical microscopy, we could see surface regions that belonged to different ferroelastic structural domains (see the inset of Fig. 8). We could also observe that the surface was bent macroscopically due to twinnings in regions where two ferroelastic domains met [23].

Apart from standard microscopy techniques, we report here that the formation of a local multi-variant state can also be detected directly using a MOKE setup. Due to the buckling effects in the multi-variant state, there is no longer only one light beam reflected off the sample in MOKE, but a set of at least two significantly split beams can be found. Beam splitting occurs in directions according to the high symmetry axis of the sample that is parallel or perpendicular to the plane of incidence. We measured field dependencies of these split beams separately by MOKE, but the results strongly varied across the sample surface (not shown). The multiplicity of the reflection originates from different twins illuminated at the same time and sharing a twinning plane laying in the plane of incidence. Scanning the sample by moving the laser beam across the surface changes not only the intensity of these reflections, but depending on the sample spot the number of reflections changed according to the local twinning structures. For example, close to the center of the sample (see the image in Fig. 8), four reflections in total were arranged into a cross as sketched in Fig. 8. The two-fold symmetry can be interpreted as reflections from two different twin domains. The second type of twin boundary can be either mutually rotated about 90 degrees in the plane of the sample or oriented perpendicularly to the surface [24], which is also in concordance with a recent paper where the structural properties were investigated by optical microscope measurements [25]. In both cases, there would be a set of two reflections symmetrically distanced from a specular reflection expected from a flat surface in the absence of twinning.

Monitoring the intensity of the specular single laser reflection observed in the austenite during cooling, we were able to precisely determine the transition temperature, where the beam intensity splits into at least two reflections. Due to a poorer focus of in the presence of multiple reflections, the intensity dropped during formation of the multi-variant martensitic phase. From this drop, we estimated an onset of the transition to be at $T_{\text{mart.}} = 330$ K, see Fig. 8. The tail during cooling indicates that the transition does not occur in one step. On the way back to the austenitic phase via heating up the sample, the intensity of the single reflection was restored and the intensity grows at $T_{aust.}$ = 338 K. Such determined temperatures of transition well correspond to the temperatures determined from magnetization measurement. This simple experiment detecting the beam splitting effect directly confirms that the change of the character of hysteresis loops originates from the transformation to the high-temperature austenitic structure at $T_{aust.}$.

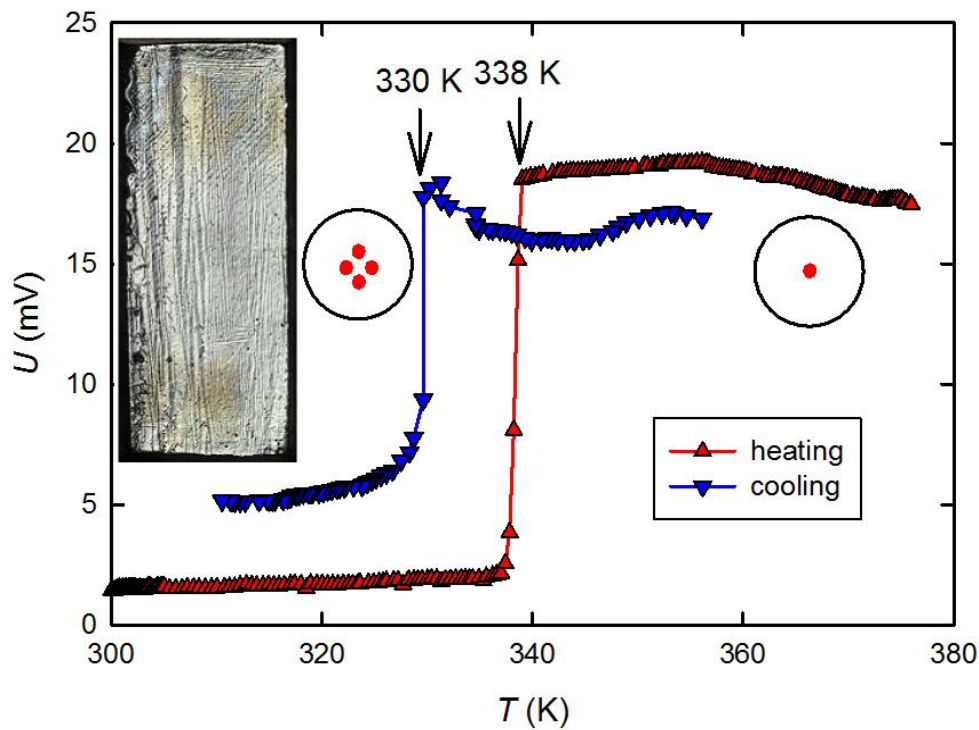

Fig. 8: Temperature scans collecting the detector intensity when focused to the specular reflection from the flat surface of the austenitic phase. Two arrows mark transformation temperatures for cooling and heating procedures. Light reflection patterns from the sample are sketched in the black circles. Multiple reflections caused by the multi-variant state occur below 330 K during cooling. On the other hand only one single reflection reappeared after heating above $T_{aust.}$ = 338 K where the sample transforms again into the austenitic structure. The microscopy image in the inset shows a sample in a multi-variant state with structural domains.

## 4. Discussion

We summarize all our MOKE experiments in Fig. 9. The Kerr rotations manifest a clear decreasing trend with increasing temperature as expected for decreasing average magnetization. Since the Kerr rotation does not change in a step-like fashion at $T_{\text{IMT, cooling}} = 213$ K and $T_{\text{IMT, heating}} = 280$ K during cooling and heating respectively, it follows that the IMT does not involve an abrupt change in the electronic band structure at least in the energy range around 1.85 eV below the Fermi level, because this would affect the magnitude of Kerr rotation accordingly. This result is in line with an adaptive concept of martensite [26] and also with reports on stoichiometric $Ni_2MnGa$, where the density of states at Fermi level does not significantly vary between different martensitic phases [27]. An intrinsic origin of the discontinuity of the Kerr values derived from *heating #2* and *heating #3* at around room temperature can be excluded by *heating #4*. The sample was heated from the lowest temperatures up to the paramagnetic state above the Curie temperature showing a continuous behavior at room temperature. The Kerr rotations among our experiments vary only due to an inhomogeneity of the sample surface. Nevertheless the magnitudes are comparable to published data [21].

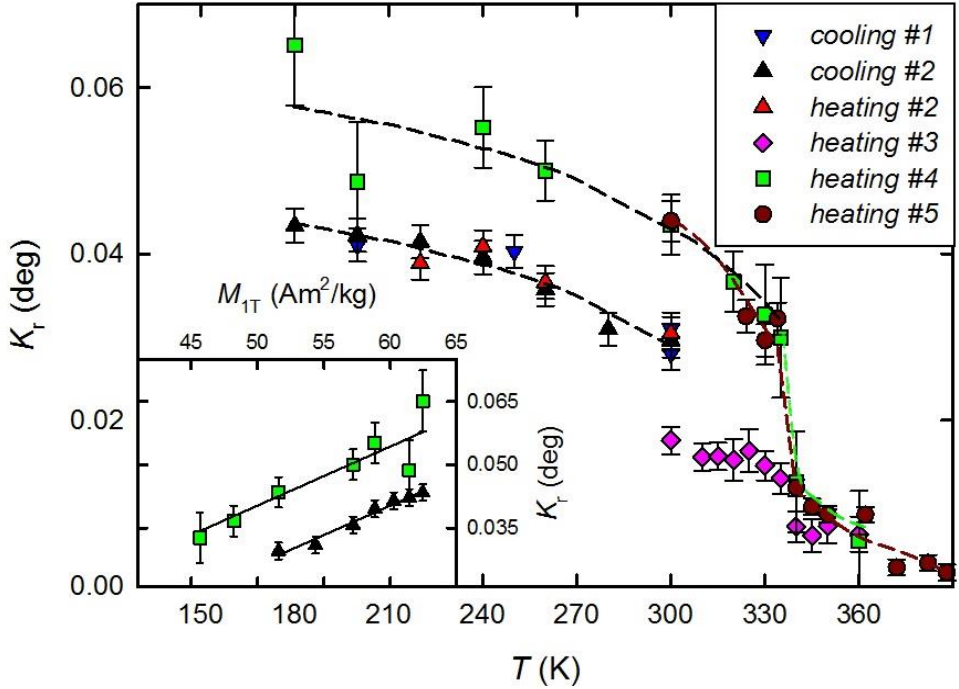

Fig. 9: A summary of the derived Kerr rotations from all our experiments. The difference in the magnitude of Kerr rotation among the experiments (especially the room temperature values) can be ascribed to a non-homogeneity of the sample surface. The inset shows linear fits of $K_r$ vs. $M$ curves. The obtained parameters served for constructing of the black dashed lines in the main figure. The dashed lines in other colors are guides to the eye.

We want to again address the fact that the overall lower values of the coercive fields in Fig. 5 found by MOKE with respect to bulk sensitive SQUID and VSM can be explained by surface pinning. Furthermore, below the IMT, the coercive fields from the bulk data increase only by approximately 40 % in comparison with the room temperature values (in contrast to a factor of two from the MOKE data) emphasizing the role of the surface pinning. We expect that the surface might have contained more defects than the volume, which could have hindered domain walls motion. This would result in enhanced magnitudes of coercive fields in the MOKE measurements.

In order to evaluate the data more quantitatively, we fitted the temperature dependencies of Kerr rotations to magnetization. We found a linear dependence of $K_r$ on $M_{1T}$ in the martensitic phase. For both *cooling #2* and *heating #4*, the data can be fitted satisfactorily with the same proportionality to $M_{1T}$ coefficient of 0.0014. We omit the fit of *heating #3* MOKE data, since this experiment covered a very limited temperature range. Furthermore, $M_{1T}$ had much steeper temperature dependence in the vicinity of $T_{aust.}$ than at lower temperatures and that deepened the discrepancy between MOKE signal and magnetization taken at 0.1 T and 1 T magnetic field, respectively.

A discontinuous jump in Kerr rotation describing an intrinsic behavior of the sample appears around $T_{aust.}$ =338 K during the transformation to the austenite. This agrees well with the published data in the literature where it is explained via changes in the electronic structure [21]. However, we have to also consider other quadratic contributions to MOKE, because in the austenite, transversal components of magnetization can be non-zero at low fields and our results on longitudinal geometry revealed a very different nature of MOKE field scans in the austenite than in the martensite. In the former case, the field scans are clearly asymmetric and remind of smeared curves, which can be found in materials exhibiting the Voigt effect [28]. As to evaluate the data in Fig. 6, we subtracted the standard s-shaped curve occurring at $T$ = 360 K from those wing-like shaped curves occurring at 340, 345, and 350 K. The remaining MOKE signals were even with respect to the zero field axis. The change of MOKE signal across the entire field range was about one half bigger than the step at 360 K pointing to a prevailing quadratic dependence of the Kerr rotation on magnetic field and consequently magnetization. But to decide which one of the quadratic terms $M_L M_T$, $M_T^2$ or $M_L^2$ plays the most important role, we would need to collect the data in different sample or magnetic field orientations [14]. Quadratic MOKE signals were also reported e.g. in a similar compound, $Co_2FeSi$ [29], or in off-stoichiometric CoMnGe thin films [30]. Around zero field, there is an overlap of the field up and field down runs, which we interpret as a formation of magnetically compensated states on the surface similarly to e.g. antiferromagnetically coupled thin films [31] or circular nanomagnets [32]. In the latter case, the nanomagnets form vortices on the surface. The same feature was found in $Ni_2MnGa$ quenched polycrystalline samples [24] and we can expect that it occurs also on the surface of the single crystal studied in this work, especially at very low fields.

As an origin of the quadratic MOKE contributions we stress, that the magnetic anisotropy in the austenitic phase is much smaller and cubic [33, 34], which - in contrast to the martensite - allows domains to orient more easily in directions other than the [001]. Moreover, owing to the very low anisotropy, magnetic domains walls are very broad and magnetization can form eddies [24]. The only significant anisotropy persisting in the austenitic phase is a shape anisotropy caused by demagnetizing fields, which can be roughly estimated considering the sample dimensions and approximating the sample by an ellipsoid [35]. Using this, we get respective ratios of the ellipsoid dimensions along the axes of the martensite: $x_a/x_c = 0.2$ and $x_b/x_c = 0.4$. At maximum magnetic fields of our MOKE setup in longitudinal geometry, the sample experiences a magnetic field of 0.13 T and the volume magnetization is 470 emu/cm$^3$ from the VSM measurements shown in Fig. 2. According to this, demagnetizing fields are approximately 0.36 T, 0.18 T and 0.05 T for the $x_a$, $x_b$, $x_c$ dimensions of the sample, respectively. If we consider the saturated magnetization to be a factor of 2 lower in the austenite at 350 K, the shape anisotropy is still larger than the maximum field in the polar geometry (0.076 T). Our polar geometry measurements (not shown) indeed did not reveal any signs of hysteresis pointing to magnetization oriented out-of-plane. Thus the magnetization should be located fully in plane and form only $M_L$ and $M_T$ contributions, otherwise it would be energetically highly unfavorable. Instead, we expect canting of magnetization by a Stoner rotation process in the austenitic state.

5. Conclusions

Our experiments have proved that the magnetic state in Ni$_2$MnGa alloys may be efficiently studied not only in the polar geometry [21], but also via the longitudinal one. For Ni$_2$MnGa sample dimensions with a high aspect ratio along the field direction, magnetic properties can be measured along the magneto-crystalline easy axis of Ni$_2$MnGa by the longitudinal MOKE giving access to local coercive field values, which are characterizing key material properties needed for future applications. The fact that longitudinal fields are also oriented in the easy axis plane defined by the shape anisotropy simplifies a modeling of magnetic reorientation processes.

This work was supported by the Czech Science Foundation Grants no. 16-00043 and OP VVV project MATFUN under Grant CZ.02.1.01/0.0/0.0/15_003/0000487, and projects of MEYS LM2015088 and LO1409. J.H. gratefully acknowledges the Purkyně Fellowship program of the AS CR.